\documentclass[pra,aps,twocolumn,superscriptaddress,longbibliography, nofootinbib, showpacs]{revtex4-1}

\usepackage[colorlinks=true, citecolor=red, urlcolor=blue ]{hyperref}
\usepackage{graphicx}
\usepackage{bm}
\usepackage{amsmath,amsfonts}
\usepackage{amsthm}
\usepackage{hyperref}
\usepackage{xcolor}
\hypersetup{
    urlcolor=magenta,
    citecolor=blue
           }

\usepackage{braket}
\theoremstyle{plain}
\usepackage{color}
\usepackage{amssymb}
\usepackage{amsthm}
\usepackage{amsfonts}
\usepackage{float}
\usepackage{tabularx}
\usepackage{graphicx}

\usepackage{mathtools}
\usepackage{esvect}
\usepackage{wrapfig}
\usepackage{amsthm}
\usepackage{verbatim}
\usepackage{bbm}
\usepackage[normalem]{ulem}

\usepackage{enumitem}
\usepackage{fmtcount}
\usepackage{booktabs}
\usepackage{csquotes}
\usepackage{epsfig}

\usepackage{tabularx}
\usepackage{graphicx}
\usepackage{color,soul}
\usepackage{amsmath}
\usepackage{braket}
\usepackage{latexsym}
\usepackage{bm}
\usepackage{graphics,epstopdf}
\usepackage{enumitem}
\usepackage{fmtcount}
\usepackage{booktabs}
\usepackage{csquotes}
\usepackage{epsfig}

\theoremstyle{plain}

\def\bea{\begin{eqnarray}}
\def\eea{\end{eqnarray}}
\def\ba{\begin{array}}
\def\ea{\end{array}}

\def\beq{\begin{equation}}
\def\eeq{\end{equation}}

\usepackage[normalem]{ulem}
\usepackage{float}
\usepackage{graphicx}  
\usepackage{dcolumn}          
\usepackage{amssymb}
\usepackage{appendix}
\usepackage{physics}   
\usepackage{mathtools}
\usepackage{esvect}
\usepackage{wrapfig}
\usepackage{amsthm}

\usepackage{verbatim}
\usepackage{bbm}

\usepackage[mathscr]{euscript}

\def\({\left(}
\def\){\right)}
\def\[{\left[}
\def\]{\right]}

\begin{document}
\pagenumbering{arabic}
\title{Effects of noise on performance of Bernstein-Vazirani algorithm}

\author{Archi Gupta}
\affiliation{Harish-Chandra Research Institute,  A CI of Homi Bhabha National
Institute, Chhatnag Road, Jhunsi, Prayagraj - 211019, India}
\affiliation{Visvesvaraya National Institute of Technology, South Ambazari Road, Nagpur, Maharashtra. 440010, India}
\author{Priya Ghosh}
\affiliation{Harish-Chandra Research Institute,  A CI of Homi Bhabha National
Institute, Chhatnag Road, Jhunsi, Prayagraj - 211019, India}
\author{Kornikar sen}
\affiliation{Departamento de F´ısica Te´orica, Universidad Complutense, 28040 Madrid, Spain}
\author{Ujjwal Sen}
\affiliation{Harish-Chandra Research Institute,  A CI of Homi Bhabha National
Institute, Chhatnag Road, Jhunsi, Prayagraj - 211019, India}

\begin{abstract}

The Bernstein-Vazirani (BV) algorithm offers exceptional accuracy in finding the hidden bit string of a function. We explore how the algorithm performs in real-world situations where noise can potentially interfere with its performance. In order to assess the impact of imperfect equipments, we introduce various forms of glassy disorders into the effect of the Hadamard gates used in the Bernstein-Vazirani circuit. We incorporated disorders of five different forms, viz., Haar-uniform with finite cutoff, spherical Gaussian, discrete circular, spherical Cauchy-Lorentz, and squeezed. We find that the effectiveness of the algorithm decreases with increasing disorder strength in all cases. Additionally, we demonstrate that as the number of bits in the secret string increases, the success probability of correctly guessing the string becomes increasingly insensitive to the type of disorder and instead depends only on the mean and spread of the disorder. We compare our results with the performance of the analogous classical algorithm in the presence of similar noise. When the length of the secret string is small or moderate, the quantum BV algorithm is found to be more efficient compared to the classical algorithm for almost all types of disorders under consideration, unless the strength of the disorder is very high and the disorder follows a discrete circular distribution. However, if we move to extremely large secret strings, the success probability of the disordered BV algorithm merges with the success probability of the disordered classical algorithm for all considered disorders having arbitrary strengths. The limit on the length of the string after which the efficiency of the quantum algorithm becomes equivalent to the classical algorithm depends on the amount of disorder and not on the type of disorder.


\end{abstract}

\maketitle

\section{Introduction}
\label{sec1}

Quantum information and computation have captured the attention of many researchers, entrepreneurs, and investors due to their potential to revolutionize different areas of industry and science~\cite{ qinfo3, qinfo2}. Quantum computers harness the fundamental principles of quantum mechanics to perform simulations~\cite{Montanaro,interference, neilson}. 
Quantum algorithms are believed to outperform their classical
cousins in terms of solving speed~\cite{algo1,algo3,algo4,algo5,algo6,algo7,algo8},
but computational errors are expected to be kept to a minimum to
realize this their potential~\cite{outperform,Montanaro}.  
Examples include Shor's factoring algorithm~\cite{shor}, Deutsch-Jozsa algorithm~\cite{Deutsch}, Grover algorithm~\cite{grover}, phase estimation algorithm~\cite{phaseest}, etc.

Bernstein and Vazirani observed that the Deutsch-Jozsa~\cite{Deutsch} algorithm could be extended
to identify a bit string hidden inside a particular type of function~\cite{bv1,bv2}. The Bernstein-Vazirani (BV) algorithm outperforms its classical counterpart for longer than one bit string without exploiting entanglement~\cite{review_paper, no-ent, Prl_no_ent, suchetana_bv}. It should be noted that the length of the hidden bit string must be more than one to experience quantum improvement. The performance of the quantum algorithm, i.e., the BV algorithm, is directly related to the quantum coherence of the initial state~\cite{suchetana_bv}.
Bernstein-Vazirani parity problem can be
experimentally solved without using entanglement~\cite{purity_exp}. Fiber-optics and various modified optical instruments have been used to implement the algorithm~\cite{fibre, puzzle, optical, ion, optical2}. The BV algorithm can be modified to find the square root of an integer number and multiplication of integer numbers~\cite{ beyond-qubit,qudit,generalize, integer}. 
A novel entanglement based quantum key distribution protocol is proposed which utilizes a modified symmetric version of BV algorithm~\cite{qkd}.

In practical situations, it is almost impossible to execute quantum operations without experiencing any noise, which may be caused by several factors such as decoherence, control errors, cross talk, readout errors, initialization errors, environmental influences, or hardware imperfections~\cite{noise3}. Therefore, it becomes crucial to thoroughly investigate and analyze the effect of noise on the performance of the quantum algorithms.

In the paper, we investigate the efficiency of BV algorithm in presence of a special type of disorder, called ``glassy"~\cite{glassy1,glassy2,glassy3,glassy4,glassy5,glassy6} disorder. In literature, it is also known as ``quenched" disorder. Glassy disorder refers to a disorder in which one or more system parameters may take random values and the observation time is much smaller than the equilibrating time of the system with the disordered parameter or parameters. We consider the case when all the Hadamard gates, used in the BV algorithm, are affected by the glassy disorder. 
To explore the performance of the disordered BV algorithm, we examine the behavior of the disorder-averaged probability of correctly determining the secret string in one query, with the strength of the noise. The probability is found to be independent of the form of the actual secret string, as far as the length is fixed and the distribution contains reflection symmetry about a certain plane. Using central limit theorem~\cite{clt_shoe,book-ch6,monte-carlo}, we analytically prove that all disorders having well-defined mean and standard deviation have the same impact on the efficiency, defined to be the probability of detecting correct string, of BV algorithm when the length of the secret string is considerably large.

The efficiency of the classical noiseless algorithm, which searches the secret string in one query, is known to be decreasing exponentially with the length of the string. To further compare the quantum and classical algorithms, we also employ glassy disorder within the gates used in the classical algorithm. We consider five different distributions of the glassy-disorder, i.e., uniform with finite cutoff, Gaussian, Cauchy-Lorentz, discrete, and squeezed. Our results indicate that when the amount of disorder, incorporated in the BV algorithm, is considerably small, the corresponding disorder-averaged probability decreases slowly with the strength of the employed disorder for a fixed length of the string. But above a threshold amount of the disorder's strength, the probability starts to decrease fast, finally reaching an almost negligible value. Then, we check the difference between disordered quantum and classical success probabilities with respect to the length of the string of bits for various types of disorder and disorder strengths.
We realize that the disordered BV algorithm is consistently more efficient than its classical counterpart in searching small or moderate-sized hidden strings when the disorder under consideration does not have very high strength. For the case of strong discrete disorder, the classical algorithm depicts better performance than the quantum one when number of bits in secret string is again comparatively small or moderate. If we increase the length of the secret string further, the discrepancy between the disordered quantum and classical algorithms decreases, finally becoming almost equal for all considered types of disorder having arbitrary strengths. The range of the string length for which the disordered BV algorithm can provide an advantage over the classical algorithm depends on the type and strength of applied disorder, i.e., the range decreases with increasing disorder strength. 

The structure of the remaining part of the paper is constructed 
as follows. In Sec.~\ref{sec2}, the BV algorithm is briefly recapitulated. In Sec.~\ref{sec3}, we discuss how the disorder is incorporated into the BV algorithm. In section~\ref{sec4}, we prove that whatever the type of disorder present in the BV algorithm, the response of the BV algorithm will be the same, provided the distribution's mean and variance is well-defined and fixed, and the number of bits is large. The different types of probability distributions of points on the surface of a unit sphere that we use for prototyping the disorder distributions in the BV algorithm are discussed in section~\ref{sec5}. We present our analytical and numerical results about the disorder-averaged probabilities of success of the BV algorithm for different disorder distributions in the Sec.~\ref{sec6}. The quantum algorithm, i.e., the BV algorithm is compared with its classical cousin in the same section. Conclusive remarks drawn from the comprehensive analysis are expounded in Sec.~\ref{sec7}.

\section{Bernstein-Vazirani Problem}
\label{sec2}
The BV algorithm is often regarded as a cousin of the Deutsch-Jozsa algorithm. In the BV algorithm, instead of determining
whether a binary function is constant or balanced, one tries to detect the hidden bit string encoded inside a function. The function maps a string of bits to a binary number. The hidden bit string is determined by measuring the function's outcomes corresponding to different inputs.

 The effectiveness of an algorithm can be defined in terms of the number
of queries required to completely solve a particular problem~\cite{query}.
The usual noiseless BV algorithm can detect all the bits of the secret string in a single query. This is  achievable due to the ability of quantum systems to execute parallel operations on a group of orthogonal states presented in a single superposition. This stands in contrast to the corresponding classical algorithm, which can only perform operations sequentially on each of the individual states.

\subsection{Detection of the secret string}
In this subsection, we briefly discuss the motive of the algorithm. In the following two parts, we will present the structure of the classical and quantum algorithms used to solve the problem.

Let there be a function of the type, 
$f: \{0,1\}^n \rightarrow \{0,1\}$. The actual operation of the function is defined as $f(x)= s \cdot x$, where `$\cdot$' denotes bit-wise multiplication modulo $2$ and $s$ is a constant secret string having $n$ number of bits. The function, $f$, is available as a black-box and the string, $s$, is unknown. The objective is to identify the correct form of $s$ with the lowest possible number of iterations of the function $f$.

\subsubsection{Classical approach}

Classically, if we try to guess a bit-string, hidden inside the black-box, without using any proper algorithm, it would be very inefficient, as a string of $n$ unknown bits can have $2^n$ forms. The best way is to try finding each of the bits of the secret string, one at a time, using some particular inputs. For example, to determine each bit of the $n$ bit secret string, $s \equiv s_1 s_2  s_3 \ldots s_n $, the inputs can be chosen in such a way that we get
\begin{center}
$f(100 \ldots 0) = s_1,$\\
$f(010 \ldots 0) = s_2,$\\
$f(001 \ldots 0) = s_3,$\\
.\\
.\\
.\\
$f(000 \ldots 1) = s_n$,\\
\end{center}
where $s_k$ denotes the value of the $k$th bit of $s$. This implies that when the function acts on a bit-string having exactly single 1, at $k$th position, it provides the value of the $k$th bit of the string, $s$, as an output. Thus, one needs to operate the function, $f$, at least $n$ times, to disclose each bit of the string having length $n$. We can think that the algorithm takes a string of $n$-bits, all of which are initialized in the state 0, as input. In the algorithm, the $k$th bit is flipped from 0 to 1, then passed inside the function $f$ to get information about the $k$th bit of the secret string. Thus, to get the complete information about the $n$-bit secret string, the bit flip operator should be used total $n$ number of times.


\subsubsection{Quantum approach}
\begin{figure}[H]
\centering
\includegraphics[scale= 1.0]{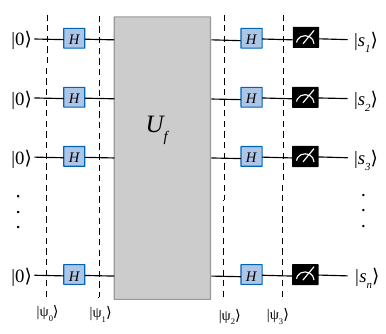}
\caption{\textbf{Schematic diagram of the quantum circuit used for the BV algorithm.} In the figure, the small blue boxes labeled with~\textit{H} represent the Hadamard gates, the light gray colored big box in between the two sets of Hadamard gates, labeled as $U_f$, is the oracle encoded with the function, $f$, and the small black colored boxes represent the measurement operators. The dashed lines depict the barriers used to separate the different states ($\ket{\psi_0}, \ket{\psi_1}, \ket{\psi_2}$, and $\ket{\psi_3}$) from each other. This version of the BV circuit is often regarded as the economical circuit of the BV algorithm, as it uses only $n$ input qubit to find an unknown $n$-bit string~\cite{review_paper}.}
\label{bvckt}
\end{figure}

Running the BV algorithm on a quantum computer has the potential to unlock the mystery of the secret string in a single attempt. The BV algorithm uses $n$ qubits as input, and provides the $n$-bit secret string $s$ as output, $\ket{s_1 s_2 s_3\ldots s_n}$. The circuit for the BV algorithm is shown in Fig. \ref{bvckt}. 

The circuit for the oracle is nicely explained in Ref.~\cite{review_paper}. All the $n$ input qubits of the algorithm are initially prepared in $\ket{0}$ as $\ket{\psi_0} = \ket{0}^{\otimes n}$. Then $n$ Hadamard gates, $H^{\otimes n}$, are applied on $\ket{\psi_0}$, which transforms $\ket{\psi_0}$ to the state $ \ket{\psi_1} =  \frac{1}{\sqrt{2^n}} \left(  \ket{0} + \ket{1} \right)^{\otimes n}$. The oracle, $U_f$, transforms a state, $\ket{x}$, as 
${{U}_{\text{f}}} \ket{x} = (-1)^{f(x)} \ket{x}$. After the action of the Hadamard gates, the oracle is performed on $\ket{\psi_1}$, and thus we get
$\ket{\psi_2} = {{U}_{\text{f}}} \ket{\psi_1}= \frac{1}{\sqrt{2^n}} \bigotimes_{k = 1}^n \left[  \ket{0} + (-1)^{s_k} \ket{1} \right]$. Again, a set of $n$ Hadamard gates is applied  on the state, $\ket{\psi_{2}}$, which transforms it to the state  $\ket{\psi_3}  = H^{\otimes n} \ket{\psi_2}   = \bigotimes_{k = 1}^{n} \ket{s_k}.$ The quantum state, $\ket{\psi_3}$, actually represents the string $s$. Therefore, by measuring each and every qubit of the state, $\ket{\psi_3}$, in the computation basis, $\{\ket{0},\ket{1}\}$, one can achieve complete information about the entire secret string $s$, using a single operation of the function, $f$.

\section{Infusion of Noise in the BV algorithm}
\label{sec3}
The purpose of this section is to analyze how the BV algorithm performs in the presence of noise. BV algorithm uses $2n$ Hadamard gates to discover $n$ bit secret string. A Hadamard gate transforms the zero coherent state $\ket{0}$ ($\ket{1}$) to a maximally coherent state $(\ket{0}+\ket{1})/\sqrt{2}$ ($(\ket{0}-\ket{1})/\sqrt{2}$) where the coherence is being measured in the computational basis, $\{\ket{0},\ket{1}\}$. However, due to the inherent fragility of quantum devices, the implementation of these quantum gates can be subjected to various sources of noise and errors. In this paper, we consider the presence of a special kind of disorder called `glassy' or `quenched' disorder in all the Hadamard gates.

Glassy disorder refers to a specific type of disorder in the physical systems that results in a random change in one or multiple parameters that becomes frozen over time. This means that the particular realization of disordered parameter does not evolve appreciably with time. Glassy disorders may arise within the Hadamard gates due to a variety of factors, such as the use of imperfect hardware, external noise, and interference. Therefore, it is quite an interesting and important task to investigate how the BV algorithm will work when all the Hadamard gates are affected by noise.

We incorporate the glassy disorder in all the 2$n$ number of Hadamard gates, involved in the BV algorithm. Let the disordered Hadamard gate, $H(\theta,\phi)$, transforms the states, $\ket{0}$ and $\ket{1}$, to $\ket{\eta(\theta,\phi)}$ and $\ket{\eta^{\perp}(\theta,\phi)}$, respectively, in the following way:
 \begin{equation}
      H(\theta,\phi) \ket{0}=\ket{\eta(\theta,\phi)} := \cos\frac{\theta}{2} \ket{0} + e^{i \phi } \sin\frac{\theta}{2} \ket{1},  \label{eq-h1}
 \end{equation}
 \begin{equation}
       H(\theta, \phi) \ket{1}= \ket{\eta^{\perp}(\theta,\phi)} :=\sin \frac{\theta}{2} \ket{0} - e^{i \phi } \cos \frac{\theta}{2} \ket{1},   \label{eq-h2}
 \end{equation}
where $(\theta,\phi)$ is the spherical polar coordinate of the point which represents the state $\ket{\eta(\theta,\phi)}$ on the surface of the Block sphere. The states $\ket{\eta(\theta,\phi)}$ and $\ket{\eta^{\perp}(\theta,\phi)}$ are certainly orthogonal to each other. The values of $\theta$ and $\phi$ depend on the type and amount of disorder and can be chosen from an appropriate probability distribution which is able to describe the specific disorder. Generally, the range of $\theta$ and $\phi$ are $[0,\pi]$ and, $[0,2 \pi)$, respectively. But depending on the distribution, the range can be shortened.

Because of the imperfections present in the Hadamard gates, the set of outputs of the measurements, performed at the end of the algorithm, might be different from the actual bits of the string. Our interest lie on the probability of obtaining the correct secret string in a single query. We call this probability as the `success probability'. {The noisy Hadamard gates, operating on the $k$-th qubit, before and after the action of the oracle, are represented by $H(\theta_k,\phi_k)$ and $H(\theta'_k,\phi'_k)$, respectively. The state, $\ket{\psi'_3}$, after the action of the disordered Hadamard gates, $H(\theta_k,\phi_k)$, followed by the oracle, $U_f$, and another set of noisy Hadamard gates, $H(\theta'_k,\phi'_k)$, on $\ket{\psi_0}$ is given by
\begin{equation*}
\ket{\psi'_3}=\bigotimes_{k=1}^nH(\theta'_k,\phi'_k)U_fH(\theta_k,\phi_k)\ket{\psi_0}.
\end{equation*} }
Using Eqs.~\eqref{eq-h1} and~\eqref{eq-h2} we get
\begin{widetext}
    \begin{equation*}
  \ket{\psi'_3}  =\bigotimes_{k=1} ^ n  \Bigg[\left( \cos \frac{\theta_k}{2} \cos\frac{\theta_k ' }{2} + (-1)^{s_k} e^{i \phi_k } \sin\frac{\theta_k }{2} \sin\frac{\theta_k ' }{2}\right) \ket{0} + e^{i \phi_k '} \left(\cos \frac{\theta_k}{2} \sin \frac{\theta_k '}{2} - (-1)^{s_k} e^{i \phi_k } \sin \frac{\theta_k}{2} \cos\frac{\theta_k ' }{2} \right) \ket{1}\Bigg]. \label{eq1}
\end{equation*}
\end{widetext}
For details, see Appendix A. In the presence of disorder, after performing the measurement in the computational basis, the probability of getting the correct string of bits as output in a single query will be 
\begin{equation}
P=\prod_{k\in B_s} \prod_{k'\in \Bar{B}_s}  |X(\theta_k,\theta'_k,\phi_k)|^2 |Y(\theta_{k'},\theta'_{k'},\phi_{k'},\phi'_{k'})|^2,\label{eq2}
\end{equation}
where
 \begin{widetext}
\begin{eqnarray*}
     X(\theta_k,\theta'_k,\phi_k)= \cos \frac{\theta_k}{2} \cos\frac{\theta_k ' }{2} + e^{i \phi_k } \sin\frac{\theta_k }{2} \sin\frac{\theta_k ' }{2},~
    Y(\theta_{k'},\theta'_{k'},\phi_{k'},\phi'_{k'})= e^{i \phi_{k'} '} \left(\cos \frac{\theta_{k'}}{2} \sin \frac{\theta_{k'} '}{2}+ e^{i \phi_{k} } \sin \frac{\theta_{k'}}{2} \cos\frac{\theta_{k'} ' }{2} \right).
\end{eqnarray*}
\end{widetext}   
Here $B_s$ and $\Bar{B}_s$ are the sets of integers representing the bit position at which the string, $s$, has 0 and 1, respectively. Let the cardinality of $B_s$ and $\Bar{B}_s$ be $n_{B_s}$ and $n_{\Bar{B}_s}$. The sum of the cardinalities, i.e., $n_{B_s}+n_{\Bar{B}_s}$, is equal to the length of the string, $n$. For example, if the actual string is $s=0110$ then $n=4$, $B_s=\{1,4\}$, $\Bar{B}_s=\{2,3\}$, and $n_{B_s}=n_{\Bar{B}_s}=2$.

Since in the case of glassy disorder the observation time of a physical quantity of a disordered system is much smaller than the equilibration time of the system with the disorder configuration, to explore the nature of that quantity we can examine the average behavior of that quantity, averaged over the entire disorder distribution. In this paper, the quantity under interest is the success probability of the algorithm, $P$. 
Let all the disordered Hadamard gates be independent and follow the same disorder distribution, defined by the probability density function, $f(\theta,\phi)$, of the spherical polar angles $(\theta,\phi)$.  Then the disorder average of $P$ in our case can be written as
\begin{widetext}
\begin{equation*}
    Q= \frac{1}{(4\pi)^{2n}} \left(\int_{\mathcal{BS}} G_1(\theta,\theta',\phi,\phi') \sin\theta \sin\theta'd\theta d\phi d\theta' d\phi' \right)^{n_{B_s}}\left(\int_{\mathcal{BS}} G_2(\theta,\theta',\phi,\phi') \sin\theta \sin\theta' d\theta d\phi d\theta' d\phi'\right)^{n_{\bar{B}_s}}, \label{eq-q}
\end{equation*}
\end{widetext}
where $G_1(\theta,\theta',\phi,\phi')=|X(\theta,\theta',\phi)|^2 f(\theta,\phi)f(\theta',\phi')$,  $G_2(\theta,\theta',\phi,\phi')=|Y(\theta,\theta',\phi,\phi')|^2 f(\theta,\phi)f(\theta',\phi')$. The symbol `$\mathcal{BS}$', written in the suffix of the integration, confirms that the range of the integration is taken over the entire Block sphere. Since $(\theta,\phi)$ are defined to be the zenith and azimuthal angles of the Bloch sphere, the ranges of $\theta$ and $\phi$ are [0,$\pi$] and [$0$,$2\pi$), respectively. It can be easily checked that if the disorder distribution has reflection symmetry about the $x-y$ plane of the Block sphere, i.e., if $f(\theta,\phi)=f(\pi-\theta,\phi)$, then 
\begin{eqnarray*}
\int_{\mathcal{BS}} G_1 \sin\theta \sin\theta' d\theta d\phi d\theta' d\phi'\\ = \int_{\mathcal{BS}} G_2 \sin\theta \sin\theta' d\theta d\phi d\theta' d\phi'.
\end{eqnarray*}
Most of the typical distributions found in nature contain this symmetry. Thus from now on we will only consider the disorder distributions which follow this symmetry. The final form of the disorder-averaged success probability of determining the $n$-bit secret string in a single query, considering this type of distribution, i.e., the distribution which has reflection symmetry about the $x-y$ plane of the Bloch sphere, reduces to 
\begin{equation*}
    Q= \frac{1}{(4\pi)^{2n}} \left(\int G_1 \sin\theta \sin\theta' d\theta d\phi d\theta' d\phi' \right)^{n}.
\end{equation*}

\section{Distinct disorders exhibiting similar response}

\label{sec4}
Our first motive is to check whether different types of disorders affect the success probability of the BV algorithm differently. To investigate this, we will invoke the central limit theorem, which we briefly outline below. 
\newline
\newline
\textbf{\textit{Central Limit Theorem:}}
The central limit theorem is a very powerful and dominating theorem in probability and statistics, as it allows us to make decisions about any large sample chosen from a distribution without even going into the details of the distribution itself~\cite{clt_shoe,book-ch6}. 
It can be stated as: The distribution of the sum of a sample of random numbers is always approximately Gaussian if the sample size is sufficiently large. The mean and the variance of the Gaussian distribution depend only on the individual random numbers' distributions' means and standard deviations, respectively.~\cite{monte-carlo}

Let $R_j$s represent the independent random variables, each of which is taken from a distribution described by the probability distribution function $F_j(\mu_j,\sigma_j)$, with a finite mean, $\mu_j$, and finite standard deviation, $\sigma_j$, where $j=1,2,3,\cdot\cdot\cdot,N$ and  $N$ denotes the population of the sample, $\{R_j\}_j$.  Here $F_j(\mu_j,\sigma_j)$ may not be the same function as $F_k(\mu_k,\sigma_k)$ for $j\neq k$. The sum of the random variables is given by $S_N=\sum_{j=1}^N R_j$. According to the central limit theorem, for $N \rightarrow \infty$, the sum $S_N$ follows a Gaussian distribution with mean $ \mu = \frac{1}{N} \sum_{j=1} ^N \mu_j$ and standard deviation $ \sigma =  (\frac{1}{N}\sum_{j=1} ^N \sigma_{j} ^2)^{1/2}$.
Note that $N$ needs to be very large unless it is taken from a Gaussian distribution, for which it can be any positive integer.

Returning to the BV algorithm, the success probability of finding the correct secret string of $n$ bits in one trial is expressed in Eq.~\eqref{eq2}. Taking natural logarithm on both sides of the equation, we get 
\begin{equation}
    \ln P = 2\sum_{k=1} ^n z_k  ,
    \label{eq-sum}
\end{equation}
where for $1\leq k\leq n_{B_s}$, $z_k=|X(\theta_k,\theta'_k,\phi_k)|$ and for  $n_{B_s} < k \leq n$, $z_k=|Y(\theta_{k-n_{B_s}},\theta'_{k-n_{B_s}},\phi_{k-n_{B_s}},\phi'_{k-n_{B_s}})|$.
Since $\theta_k,~\phi_k,~\theta_k',$ and $ \phi_k'$
are independently distributed, the variables $z_k$, used in  Eq.~\eqref{eq-sum}, are also independent from each other, for all $k$. Using the central limit theorem, we can state that if each of the $n$ independent random variables, $z_k$, follow a distribution with finite mean, $\widetilde{\mu}_k$, and finite standard deviation, $\widetilde{\sigma}_k$, then $\ln \text{P}$ will follow a Gaussian distribution of the following form
\begin{equation*}
    \ln P \sim  \mathcal{ N} (n \mu^*, \sqrt{n} \sigma^*), \label{eq-N}
\end{equation*}
for large $n$. Here, $n \mu^*$ and $\sqrt{n} \sigma^*$ denote $\sum_k \widetilde{\mu}_k$ and $(\sum_k \widetilde{\sigma}_k^2)^{1/2}$. Hence, we can conclude that the success probability, $P$, will follow a log-normal distribution when number of independent variables $n \rightarrow \infty$. Therefore, the mean of $P$ approaches to
    $ \text{exp}(n \mu^* + n \sigma^{*2} /2),$
 which only depends on the mean and standard deviation of the corresponding distribution of the $n$ number of independent random variables, $z_k$. This implies that in case of a sufficiently long secret string, the disorder-averaged probability, $Q$, of successfully getting the correct secret string does not depend on the minute details of the disorder distribution of the Hadamard gates, but is equal to $ \text{exp}(n \mu^* + n \sigma^{*2} /2)$.


However, it should be noted that the central limit theorem holds only when the probability distribution functions describing the random variables have well-defined means and standard deviations, which is not true in the case of Cauchy-Lorentz distribution. Thus, in Sec. \ref{secD}, we will see that incorporation of Cauchy-Lorentz disorder in the Hadamard gate's parameters have distinct effect from the other distributions, even when the length of the secret string is much larger. 
 
\section{Typical probability Distributions on sphere}
\label{sec5}
From here on, we denote a point on the Bloch sphere as $(1,\theta,\phi)$, where $\theta$ and $\phi$ represent the zenith and azimuthal angles of the point defined according to the usual convention.

Ideally, the Hadamard gate transforms $\ket{0}$ to $(\ket{0}+\ket{1})/\sqrt{2}$, but because of the presence of disorder, the Hadamard gate may transform it to a different state, represented by a point $(1,\theta,\phi)$ on the Block sphere. In this section, we briefly discuss the different types of distributions of $\theta$ and $\phi$, considered for incorporation of glassy disorder, which we will use further to study the efficiency of the BV algorithm. The considered disorder distributions are Haar-uniform with a finite cutoff, spherical-Gaussian, spherical Cauchy-Lorentz, circular and squeezed. Since, for the noiseless Hadamard gate, $ H \ket{0} = (\ket{0}+\ket{1})/\sqrt{2}$, we fix the direction of the mean of all disorder distributions along the point $(1,\pi/2, 0)$, i.e., along the positive $x$-axis. We use the standard deviations of the distributions to parameterize the strength of the corresponding disorder. For any disorder distribution on the Bloch sphere, the variance is given by the average of the square of the distance of random points from the point on the surface of the Bloch sphere in the direction of the mean. Here by `distance' we mean the length of the shortest curved line joining the two points over the surface of the sphere. For our convenience, we determine the variance by taking the mean to be along $z$-axis, which is not inappropriate because the variance is expected to be independent of the direction of the mean. Then the variance can be calculated using the following equation:
\begin{equation}
    \sigma_{QX}^2 = \frac {\int \int \theta^2 f(\theta,\phi)  \sin \theta d \theta  d \phi}{\int \int f(\theta,\phi) \sin \theta d \theta  d \phi}. \label{eq-var}
\end{equation}
Here, $f(\theta,\phi)$ is the probability distribution function of the disorder having mean along $z$-axis.

\subsection{Haar-uniform distribution with finite cut-off}
\label{secA}
We first consider the disorder to be such that the points $(1, \theta, \phi)$ are uniformly distributed on the surface of the Bloch sphere around the mean $(1,\pi/2,0)$, within a definite ``cut-off". The cut-off is defined in terms of a plane, parallel to the $y-z$ plane which cuts the Bloch sphere, giving birth to two semi-spherical caps. The points, $(1, \theta, \phi)$, are uniformly distributed on the curved surface of the spherical cap, which is situated along the positive $x$ direction of the plane. Let $d_U$ be the angle (in radians)
between an arbitrary point of the perimeter of the spherical cap and (1,$\pi$/2,0) made at the center of the sphere. The variance of the uniform distribution is found using Eq.~(\ref{eq-var}), which on further simplification reduces to be a function of $d_U$ of the following form:
\begin{eqnarray*}
     \sigma_{QU}^2 &=& \frac{\int_0 ^{d_U} \theta^2 \sin \theta {d} \theta}{\int_0 ^{d_U} \sin \theta {d} \theta} \\&=& \frac{\cos{d_U}(d_U^2-1) - 2({d_U} \sin {d_U}-1)}{\cos {d_U} - 1}.\label{eq-var-uni}
 \end{eqnarray*}
The parameter $d_U$ can vary from $0$ to $\pi$. For $d_U=0$, $\sigma_{QU}=0$. The standard deviation increases with the angle, $d_U$. When $d_U=\pi/2$ (represents distribution of points over half of the Bloch sphere), $\sigma_{QU}=1.07$ and when $d_U=\pi$ (represents distribution of points over the entire sphere), $\sigma_{QU}=1.71$. 
 
\subsection{Spherical Gaussian distribution}
Next, we consider the von Mises-Fisher distribution, which is usually referred to as the Gaussian distribution of points on the surface of a sphere~\cite{sph_nor}. The noisy Hadamard gates are chosen in such a way that the points on the surface of the Bloch sphere representing the states, $\eta(\theta,\phi)$, be spherical–normally distributed on the surface of the Bloch sphere with mean fixed at $(1, \pi/2, 0)$. The probability density function of $(\theta,\phi)$ for spherical Gaussian distribution can be defined in terms of two parameters, $\kappa_G$ and $\boldsymbol{\mu}_G$, as shown below:
\begin{equation*}
    f_{QG}(\theta,\phi) = N_{QG} \exp(\kappa_G \boldsymbol{x}^T \cdot \boldsymbol{\mu}_G).
\end{equation*}
Here, $N_{QG}=\kappa_G/ (4 \pi \sinh \kappa_G)$ is the normalization constant,  $\kappa_G$ represents the concentration parameter with the condition $\kappa_G \geq 0$, $\boldsymbol{x}$ denotes the Cartesian coordinate, $(\sin \theta \cos \phi, \sin \theta \sin \phi, \cos \theta )$, of the randomly generated point, on the surface of the sphere and $\boldsymbol{\mu}_G$ is the unit vector along the direction of the mean. As $\kappa_G$ increases, the distribution concentrates around the mean direction, $\boldsymbol{\mu}_G$. For
 $\kappa \rightarrow 0$, the points get distributed almost uniformly over the surface of the entire Bloch sphere.
To calculate the variance of this distribution, we take the mean along $z$ direction. Thus, we find $\boldsymbol{x}^T \cdot \boldsymbol{\mu}_G = \cos \theta$. Then the variance of the distribution reduces to
\begin{equation*}
    \sigma_{QG}^2 =\int_0 ^\pi  \frac{\theta^2 \kappa \exp(\kappa \cos \theta)  \sin \theta d \theta}{2 \sinh \kappa} . \label{eq-var-sph}
\end{equation*}

Let us now discuss the generation of the random points around the mean, $(1,\pi/2,0)$. To select points from this distribution, we first randomly create points around $(1,0,0)$ by computing the following cumulative distribution function
\begin{equation*}
    F(\theta)=\int_0 ^\theta f_{QG}(\theta,\phi) \sin \theta d \theta d \phi = \frac{e^{\kappa_G} e^{\kappa_G \cos \theta}}{2 \sinh \kappa_G}.      
\end{equation*}
Since $\theta \in [0,\pi]$, we get  $F(\theta) \in [0,1]$. If we consider $A_G=F(\theta)$, then $\theta$ can be represented as a function of $A_G$, that is
\begin{equation}
    \theta = \cos ^{-1} \left[\frac{1}{\kappa_G} \ln (e^{\kappa_G} - 2 A_G \sinh \kappa_G) \right].\label{eq3}
\end{equation}
By choosing $A_G$ and $\phi$ independently from uniform distributions within the respective ranges, i.e., $[0,1]$ and $[0,2 \pi)$, we obtain ($\theta$,$\phi$) that follows the spherical-Gaussian distribution, where $\theta$ is determined using Eq.~\eqref{eq3}. After obtaining these random points on the surface of the Bloch sphere around the mean along the z-axis, we just rotate them accordingly to get random points which follow the spherical Gaussian distribution around the desired mean, $(1,\pi/2,0)$.


 \subsection{Spherical Cauchy-Lorentz distribution}
When talking about typical probability distributions found in nature, the next distribution that comes to mind is the Cauchy-Lorentz distribution. Though at first the distribution may seem similar to the Gaussian distribution, it has a strangely different characteristic in the sense that the distribution does not have a finite mean. The Cauchy-Lorentz distribution can also be generalized on a unit sphere using two parameters, $\rho_C$ and $\mu_C$, as shown below:
\begin{equation*}
    f_{QC}(\theta,\phi) = N_{QC} (1+ \rho_C^2 - 2 \rho \boldsymbol{x}^T \boldsymbol{\mu}_C)^{-2} ,
\end{equation*}
where $N_{QC}= (1 - \rho_C^2)^2 / 4 \pi$ is the normalization constant, $\rho_C \in [0,1)$ acts as a concentration parameter, and $\boldsymbol{\mu}_C$ represents direction of mode of the distribution. If we consider the mode to be along $z$-axis, the distribution reduces to a simpler form given by
\begin{equation*}
     f_{QC}(\theta,\phi) = N_{QC} (1+ \rho_C^2 - 2 \rho_C \cos\theta)^{-2}.\nonumber
\end{equation*}
Thus, for convenience, to calculate the variance of this distribution, the mode is taken around $z$-axis. The variance is found using Eq. (\ref{eq-var}) and is given by
\begin{equation}
    \sigma_{QC}^2 = \int_0 ^\pi \frac{(1- \rho_C^2)^2 \theta^2 \sin \theta d \theta}{2 (1 + \rho_C^2 - 2 \rho_C \cos \theta)^2} .~\label{eq-var-cauchy}
\end{equation}
From Eq.~\eqref{eq-var-cauchy}, we find that as $\rho_C \rightarrow 1$, $\sigma_{QC} \rightarrow 0$ and distribution gets concentrated to a point along the mode direction, while when $\rho_C \rightarrow 0$, $\sigma_{QC} \rightarrow 1.71$ and the disorder becomes approximately uniformly distributed over the surface of the whole Bloch sphere~\cite{cauchy}. 

To find the random points which follow the spherical Cauchy-Lorentz distribution, we go through similar calculations as was done for spherical-Gaussian distribution. First, we obtain $\theta$ as a function of the cumulative distribution function, $A_C$, of Cauchy-Lorentz distribution, using the following relation
\begin{equation}
     \theta = \cos^{-1} \left[ \frac{1- \rho_C^2}{2 \rho_C} \left( 1- \frac{1- \rho_C^2}{(1+ \rho_C)^2 - 4 \rho_C A_C} \right) \right] . \label{eq4}
\end{equation}
$A_C$ is then chosen randomly from uniform distribution within the range $[0,1]$ and correspondingly $\theta$ is obtained using Eq.~\eqref{eq4}. We select $\phi$ uniformly from its complete range, i.e., $[0,2 \pi)$. The pair of random angles, $(\theta,\phi)$, follows the spherical Cauchy-Lorentz distribution with mode along the $z$-axis. Finally, we rotate the points about $y-$axis to get the desired distribution around the $x$-axis.


\subsection{Discrete circular distribution} 
The circular distribution, also known as ``polar distribution", is the uniform distribution of random values of angles in the range $[0, 2\pi)$. Specifically, the discrete circular distribution around the mean $(1,\pi/2,0)$ represents uniformly distributed points on the perimeter of a circle of fixed radius. We consider the disordered Hadamard gate to transform the state $\ket{0}$ to a random state which can be represented by a point on the perimeter of a circle, parallel to the $y-z$ plane, drawn symmetrically around the mean $(1,\pi/2,0)$, on the surface of the Bloch sphere. Such circles can be defined using a parameter, $d_D$, which is the angle made by the points on the circle with the line joining the center of the circle and the Bloch sphere at the center of the sphere. 
For a particular discrete circular distribution this circle is fixed and the probability of transforming $\ket{0}$ to a point on the circle is uniformly distributed and on any point outside the circle is 0.
To calculate the strength of this circular distribution, we imagine the random points to be distributed around $z$-axis on the perimeter of a circle, parallel to the $x-y$ plane, on the surface of the Bloch sphere. Using Eq. (\ref{eq-var}), we get the standard deviation of the distribution for fixed $d_D$, i.e.,
\begin{equation}
    \sigma_{QD} = {d_D} .\nonumber
\end{equation}

The points on the circular distribution on the sphere for different noise strengths can be generated by fixing $\theta$ at $d_D$ and randomly choosing $\phi$ within the whole range, $[0,2\pi)$. The resulting points will have mean along the $z$ direction. Thus, we have to just rotate these points properly to get a discrete circular distribution having mean along $x$-axis. Since for a particular discrete distribution, one parameter, $\theta$, of spherical polar coordinate remains fixed and takes distinct discrete values for different distributions, we refer the distribution as discrete. 

\subsection{Squeezed distribution}
Till now, all the considered distributions had rotational symmetry with respect to the mean direction.
We are now going to discuss a distribution which is not rotationally symmetric about the mean, $(1,\pi/2,0)$. In particular, the points are again considered to be distributed uniformly on the surface of the sphere, but the boundary of the distributed point is not circular but is squeezed either along $y$- or $z$-axis.
{To understand this more precisely, imagine a cylinder, with elliptical cross-section and axis parallel to $x$-axis, passing through the Bloch sphere. The intersection of the cylinder and Bloch sphere represents the boundary for the squeezed distribution.} Specifically, the projection of the uniformly distributed points on a plane parallel to the $y-z$ plane is restricted to be inside a pre-defined ellipse. We characterize the distribution in terms of the parameters representing that ellipse. Let the lengths of the two axes of the ellipse be $2a$ (parallel to $y$ axis) and $2b$ (parallel to $z$ axis). The area of the ellipse, $D= \pi ab$, depicts the spread of the distribution. The ratio, $r=a/b$, can be regarded as the degree of squeezing. Note that, $r=1$ mimics the case of Haar-uniform distribution with circular boundary parameterized using the angle $d_U=\sin^{-1}(D/\pi)$, defined in Sec.~\ref{secA}. The optimal range of $r$, i.e., $[r_{min},r_{max}]$, for a fixed $D$, can be found as:
\begin{itemize}
    \item $r \leq 1,  ~a_{min} = D/\pi, ~b_{max}=1 \implies r_{min}= D/ \pi$ ,
       
        \item $r \geq 1,  ~a_{max} = 1, ~b_{min}=D/\pi  \implies r_{max}= \pi / D$ .
\end{itemize}
We can choose a particular value of $D$ and then vary $r$ within the appropriate range to obtain different strengths of the disorder, $\sigma_{QS}$. 
Similar to the other distributions, $\sigma_{QS}$ can also be found using Eq.~\eqref{eq-var}.
 $\sigma_{QS}$ is found to attain approximately the same value for $r$ and $1/r$ with a minimum at $r=1$.

    \label{fig-squ-sig-vs-r}

\section{ Impact of disorder on the success probability}
\label{sec6}

In this section, we will consider each of the disorders mentioned in the previous section, and explore their effects on success probability of classical and BV algorithm. In particular, we will numerically choose $2n$ pairs of $(\theta,\phi)$ independently from the distributions and evaluate the corresponding value of the probability, $P$ using Eq.~\eqref{eq2}. This process can be repeated multiple times for different realizations of $\theta$ and $\phi$ and the corresponding average value of $P$ can be calculated to determine the disorder-averaged success probability, $Q_X$, for each disorder, $X$ which we denote as $Q_X$. For convergence, we need to find the average over a large set of $P$ to determine $Q_X$. We consider convergence up to three significant figures. 

When we will discuss the impact of specific distribution on the BV and classical algorithms, i.e., uniform, Gaussian, Cauchy-Lorentz, discrete, squeezed distributions we will denote them, respectively, by using the notation $U$, $G$, $C$, $D$, and $S$, in place of $X$. 

\subsection{Disordered classical algorithm}
Our main objective is to investigate whether the presence of disorder affects the superiority of the quantum BV algorithm over the classical approach. But in classical world, we don't have Hadamard gates. Thus, to bring the two algorithms on equal footing, we will incorporate disorder in the bit-flip operators used in the classical algorithm. 

The action of a noisy bit-flip operator can have either of the two outcomes `flip' or `not flip'. Let us assign a probability, $p$, for the event of not flip and correspondingly a probability, $1-p$, for flip. Ideally, $p=0$, implies that the state will surely flip, which is exactly what we expect. But because of the presence of disorder, $p$ is considered to be taking random values within the range $p \in [0,1]$. Since in this case, the disorderless value of $p$ is 0, we will quantify the strength of the disorder as the square root of the average squared distance of the randomly generated points, $p$, from its ideal value, 0. Mathematically, the strength of the disorder can be expressed as
\begin{equation}
    \sigma_{CX}=\int_0^1 p^2 f_{CX}(p)dp, \label{eq6}
\end{equation}
where $f_{CX}(p)$ is the normalized probability density function of the random variable $p$ representing the disorder $X$.

In the classical case, in absence of disorder, we need at least $n$ queries to correctly determine the entire secret string of $n$ bits. But since we require only one query in the BV algorithm, to compare it with its classical cousin, we consider the same for the latter. But with one query, even in the ordered classical case, we will be able to find only one bit of information with $100\%$ accuracy. Therefore, we can just arbitrarily guess about the other $n-1$ bits, having $2^{n-1}$ number of possible guesses. Hence, the success probability in the noiseless classical case with one query is $1/2^{n-1}$. The disorder-averaged success probability of a correct guess for classical BV algorithm about the entire string in one query would be $C_X=\frac{\text{disorder-averaged of }(1-p)}{2^{n-1}}$, i.e., $C_X=\frac{1-m_X}{2^{n-1}}$ where $m_X$ is the mean of the distribution of the disorder, $X$.

Whatever be the type of the disorder, the range of $p$ is always between $p \in [0,1]$ and thus the maximum strength of the disorder would definitely be less than equals to $1$. On the other hand, the strength of the disorder, defined over the surface of the Bloch sphere, can be as high as $\pi$, because in the case of the BV algorithm we calculate the strength in terms of the distance of the distributed points from $(1,\pi/2,0)$. Therefore, it is unfair to compare the disordered quantum and classical algorithms in terms of $\sigma_{QX}$ and $\sigma_{CX}$ as defined in Eqs.~\eqref{eq-var} and~\eqref{eq6}. Therefore, to compare them, we first fix the type of the disorder, say $X$, and then determine the maximum strength achievable by the disorder, separately, for the quantum and classical cases. Say these maximum values are $\sigma^{max}_{QX}$ (for quantum) and $\sigma^{max}_{CX}$ (for classical). Then we scale the disorder's strength as  $\Bar{\sigma}_X=\sigma_{CX}/\sigma^{max}_{CX}$ (for the classical algorithm) and $\Bar{\sigma}_X=\sigma_{QX}/\sigma^{max}_{QX}$ (for the quantum algorithm), to define a new quantifier of the disorder. We call it the ``scaled" strength of the disorder. Here $\sigma_{QX}$ and $\sigma_{CX}$ are noise strengths of the disorders, present in the bit-flip operator and the Hadamard gate, respectively. Whenever we compare the performance of the two algorithms having ``same" amount of disorder, we define it in terms of $\Bar{\sigma}_X$. It can be noticed that the range of $\Bar{\sigma}_X$, for both quantum and classical algorithms, lies between 0 and 1.

In the preceding subsections, we evaluate the disorder-averaged success probability of the quantum BV algorithm and compare it with the success probability of the corresponding classical algorithm, in presence of different classes of disorder. 

\subsection{Uniform disorder}
Let us now examine the BV algorithm in presence of Haar uniform disorder within a finite cut-off. In the first part (Sec. \ref{secB1}), we discuss two special cases, i.e., distribution over the full sphere and half of the sphere. After that, in Sec. \ref{secx}, we determine the disorder-averaged success probability, $Q_U$, of BV algorithm for the complete range of $\sigma_{QU}$.

\subsubsection{Two special cases : Full and Half sphere}
\label{secB1}
We first consider the points $(1, \theta, \phi)$ to be uniformly distributed over the full sphere. Thus, in this case, $d_U=\pi$ and $\sigma_{QU}=1.71$. The disorder-averaged of success probability, $Q_U$, of determining the $n$ bit secret string encoded inside the oracle using the quantum~\textcolor{teal}{[BV]} algorithm, is found to be $(\frac{1}{2})^n$.
Next, we fix $d_U$ in such a way that the corresponding points be uniformly distributed over half of the sphere. In this case, $d_U=\pi/2$ and $\sigma_{QU}=1.07$. The corresponding quenched probability is found to be $Q_U=(\frac{1}{2}  +  \frac{\pi}{16})^n$. 

We use numerical method to obtain disorder-averaged success probability of BV algorithm for other values of $d_U$.
The corresponding results are discussed in the following part.

\subsubsection{Uniform disorder of arbitrary strength}
\label{secx}
Let us now consider the more general case, where $\sigma_{QU}$ takes arbitrary values. We want to see the variation of disorder-averaged success probability, $Q_U$, in a single query, in presence of Haar-uniform disorder by varying the disorder strength, $\sigma_{QU}$. The parameter $d_U$ can be varied from $0$ to $\pi$, where $d_U=0$ represents the noiseless case ($\sigma^{min}_{QU}=0$) and $d_U=\pi$ represents the case when disorder is distributed uniformly over the whole sphere, providing the maximum strength ($\sigma^{max}_{QU}=1.71$).

To compare the situation with the classical algorithm, we consider $p$ to be uniformly distributed between 0 and $v_U$ for $v_U\in[0,1]$. So, the mean of the distribution is $m_U=v_U/2$. The disorder-averaged success probability of correctly finding any one bit of the secret string through one query is $1-m_U$. For $n$ bit secret string, the other $n-1$ bits can just be guessed and the total probability of correct guess would be $C_U=(1-m_{U})/2^{n-1}$ in a single query. The strength of this disorder is calculated using Eq.~\eqref{eq6} and found to be $\sigma_{CU}=(v_U)/\sqrt{3}$. Thus, the maximum strength of this disorder is $\sigma^{max}_{CU}=1/\sqrt{3}$, that is, when $v_U$ is 1. Therefore, we can express $C_U$ as a function of $\Bar{\sigma}_U=\sigma_{CU}/\sigma^{max}_{CU}$ in the following way:
$$C_U=1-\frac{\Bar{\sigma}_U}{2}.$$

\begin{figure}[H]
    \centering
    \includegraphics[width=8.5cm]{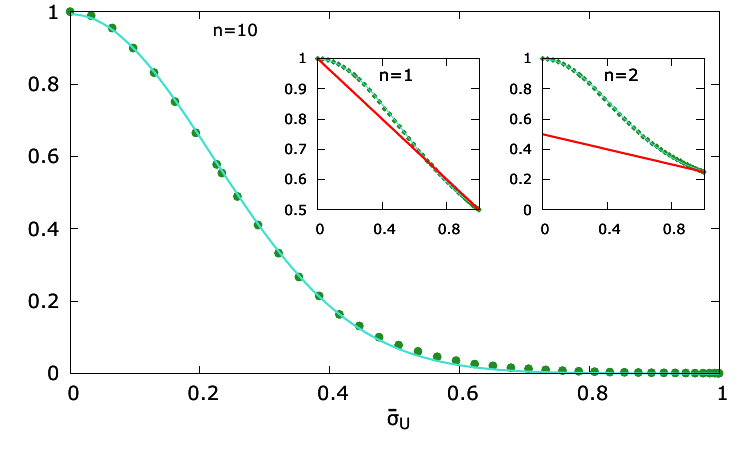}
    \caption{\textbf{Response of the classical and BV algorithms to uniform disorder}. We plot the disordered average success probabilities, $Q_U$ and $C_U$, along the vertical axis, with respect to the disorder strength ($\Bar{\sigma}_{U}$), presented along the horizontal axis, in the insets, for $n=1$ and $n=2$. For $n=10$, $C_U$ becomes almost zero. Thus for $n=10$ we show the behavior of $Q_U$ only. The green and red dots denote disorder-averaged success probabilities of the quantum algorithm and its classical counterpart, respectively. For $n=1$, though there exists a region for high disorder strength where  difference between the performance of the two algorithm is small, the classical algorithm can never overcome the quantum one. The advantage of quantum algorithm becomes more prominent for moderate-sized secret string, say $n=10$. The behavior of $Q_U$ can be fitted using a Gaussian function. The fitted curve is shown using a sky-blue line for the $n=10$ case. All the axes are dimensionless.} \label{fig:uniform-fit}
\end{figure}

We highlight the comparison between success probabilities of quantum and classical algorithms in presence of uniform disorder in Fig. \ref{fig:uniform-fit}. In the figure, we plot $Q_U$ and $C_U$ with respect to $\Bar{\sigma}_U=\sigma_{QU}/\sigma^{\max}_{QU}$ and $\Bar{\sigma}_U=\sigma_{CU}/\sigma^{\max}_{CU}$ for $n=1$ and $n=2$.
It is clear from the plots of Fig. \ref{fig:uniform-fit} that though the disorder-averaged probabilities corresponding to quantum and classical algorithms become almost the same for very high disorder, there is a significant gap in between the probabilities for moderate and weak disorders, the quantum probability being higher than the classical one. This advantage of quantum algorithm over its classical counter part reduces to a negligible amount for extremely large secret strings. See Sec.~\ref{al} for more details. For $n=1$ and very high amount of disorder, the figure indicates that the classical algorithm has slightly higher probability of successfully detecting the string than the quantum algorithm in single query. But the difference of $Q_U$ and $C_U$ in that disorder regime is negligibly small. Since $C_U$ reduces exponentially with $n$, the gap, $Q_U-C_U$, becomes non-negative in the whole range of disorder strength at the moment we move to $n>1$. $Q_U-C_U$ increases when we go from $n=1$ to 2 or 10. Since $C_U$ becomes almost zero for $n=10$, we only present the behavior of $Q_U$, and not of $C_U$, in the figure for $n=10$. Independent of the value of $n$, the relationship between $Q_U$ and $\Bar{\sigma}_U$ displays a concave shape for weak disorder, while for strong disorder, it takes on a convex form. The behavior of $Q_U$ can be nicely fitted using a Gaussian function of the form 
$$Q_U = a_U \exp(-b_U {\Bar{\sigma}_{U}}^2)+e_U,$$ where the fitting parameters, $a_U$, $b_U$, and $e_U$ depend on $n$. Values of $a_U$, $b_U$, and $e_U$ can be found using the least square method. For example, for $n=1$, the parameters $a_U=0.59 \pm 0.0024$, $b_U=1.8 \pm 0.018$, and $e_U=0.41 \pm 0.0028$, are found with error $0.0023$ whereas for $n=2$, the values are $a_U = 0.79 \pm 0.0027$, $b_U = 2.6 \pm 0.027$, $e_U = 0.19 \pm 0.0029$ with an error of $0.0047$.  Similarly, for $n=10$, the parameters become $a_U=0.99 \pm 0.0028$, $b_U=10 \pm 0.068$, $e_U = 0$ with an error of $0.0059$. 
The numbers written after all the $\pm$ symbols denote the $95\%$ confidence interval of the fitting.

\subsection{Gaussian disorder}
Next, we study the variation in success probabilities of the algorithms in presence of Gaussian disorder. For the BV algorithm, the spherical-Gaussian disorder is considered to be present in all the Hadamard gates. We examine the behavior of the success probability of the noisy algorithm with the disorder strength, $\Bar{\sigma}_G=\sigma_{QG}/\sigma^{\max}_{QG}$, where $\sigma^{\max}_{QG}=1.71$, using numerical simulations.  

For the classical case, since the range of disordered $p$ is between 0 and 1, we take the half Gaussian function to describe analogous disorder in the bit-flip operation. The half Gaussian function is given by
\begin{equation}
    f_G(p) =N_{CG} \exp(- p^2 / 2 v_G^2)
    \label{eq-sph-normal},
\end{equation}
where $N_{CG}=\sqrt{\frac{2}{\pi v_G^2}}\frac{1}{\text{erf}(1/{\sqrt{2} v_G})}$ is the normalization constant, obtained by considering the range of $p$ as [0,1]. $v_G$ is the parameter representing the spread of the disorder. `erf' represents the error function.  The mean of this distribution is given by
\begin{equation}
    m_G = N_{CG} v_G^2 \left[1 - \exp\left(-1/2 v_G^2\right)\right].\nonumber
\end{equation}
Therefore, when the length of the string is $n$, the disorder-averaged probability of successfully guessing the string using one attempt in the classical algorithm is $C_G=(1-m_G)/2^{n-1}$. The strength of the disorder described by the distribution mentioned in Eq.~\eqref{eq-sph-normal}, is
\begin{equation}
    \sigma_{CG} = v_G \sqrt{1- N_{CG} \exp(-1/ 2 v_G^2)}.\nonumber
\end{equation}
The maximum strength of this distribution is $\sigma^{max}_{CG}=1/\sqrt{3}$, i.e., when it mimics the uniform distribution over the entire range, [0,1].
\begin{figure}[H]
    \centering
    \includegraphics[width=8.5cm]{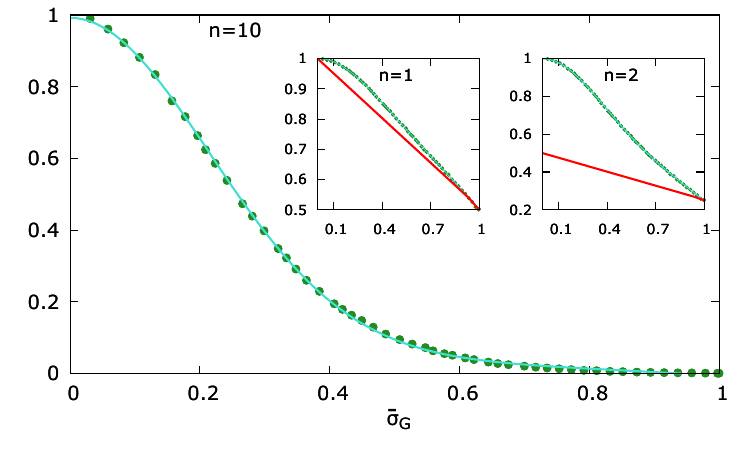}
    \caption{\textbf{Performance of the quantum and classical algorithms in the presence of Gaussian disorder}. The disordered average probability, $Q_G$, of successfully guessing the string in one query using the quantum algorithms is plotted along the vertical axis with respect to the scaled strength, $\bar{\sigma}_G$, of the applied disorder shown along the horizontal axis using green dots taking the length of the string, $n=10$. The insets show behaviors of $Q_G$ (green dot) and $C_G$ (red line) with $\bar{\sigma}_G$ for $n=1$ and $n=2$. Here the disorder under consideration is Gaussian. The curve fitted using the function expressed in Eq.~\eqref{eq5} is shown using a sky-blue line. All the axes are dimensionless. }
    \label{fig:gaussian-fit}
\end{figure}
In Fig. \ref{fig:gaussian-fit}, we plot disorder-averaged probabilities, $Q_G$ and $C_G$, as a function of the scaled disorder strength $\Bar{\sigma}_G$. One can observe from the figure that though for $n=1$ there exists a small finite range of $\Bar{\sigma}_G$ for which the classical algorithm provides almost the same efficiency as the BV algorithm, the quantum algorithm overpowers its classical counterpart, for entire disorder range, except for very high $\Bar{\sigma}_G$, as soon as one moves to $n>1$.
The classical algorithm's efficiency becomes equal to the order of $10^{-3}$ or less for the entire range of disorder strength, even in the absence of disorder, for $n>9$. The figure indicates that the quantum algorithm can perform in a more efficient way than the classical algorithm for moderate-sized bit strings if the amount of disorder is considerably small. Nevertheless, if the length of the string is increased further, keeping the disorder fixed the advantage of the quantum algorithm decreases, finally becoming equivalent to the classical algorithm for all disorder strength. More details about this can be found in Sec.~\ref{al}. The nature of the success probability of the BV algorithm has a concave to convex transition with respect to the strength, $\Bar{\sigma}_G$. The function, $Q_G$ can be fitted using a function of the following form:
\begin{equation}
    Q_G = a_G \exp (- b_G \Bar{\sigma}_{G}^2) + c_G \Bar{\sigma}_{G}^2 + e_G .
    \label{eq5}
    \end{equation}
Values of the fitting parameters, $a_G$, $b_G$, $c_G$, and $e_G$ depends on the length of the bit string and are provided in the Appendix B.




\subsection{Cauchy-Lorentz disorder}
\label{secD}
\begin{figure}[H]
    \centering
    \includegraphics[width=8.5cm]{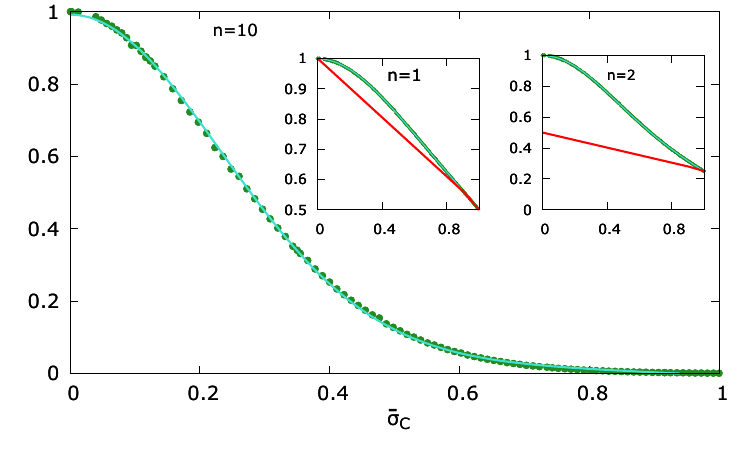}
    \caption{\textbf{Illustration of success probabilities of Cauchy-Lorentz disordered quantum and classical algorithms}. We plot the disordered average success probability, $Q_C$, along the vertical axis with respect to the disorder strength, $\Bar{\sigma}_C$, presented along the horizontal axis, considering the length of the string to be $n=10$. To compare the quantum algorithm with its classical version, in the insets, we plot $Q_C$ as well as $C_C$ along the vertical axis against the same quantity, $\Bar{\sigma}_C$, for $n=1$ and 2. The points $(\Bar{\sigma}_C, Q_C)$ and  $(\Bar{\sigma}_C, C_C)$ are shown using green and red colors, respectively. We depict the curve, fitted using the function given in Eq.~\eqref{eq5}, using a sky-blue line, for $n=10$. The axes are considered to be dimensionless.}
    \label{fig:cauchy-fit}
\end{figure}

 Let us now investigate how the performance of the classical and BV algorithms are impacted by the presence of Cauchy-Lorentz disorder.
  For the quantum algorithm, we find the behavior of the disorder-averaged success probability, $Q_C$, by varying disorder strengths, $\sigma_{QC}$. The maximum value of the disorder strength in the quantum case is $\sigma^{max}_{QC}= 1.71$. In Fig. \ref{fig:cauchy-fit}, we plot $Q_C$ as a function of $\Bar{\sigma}_C=\sigma_{QC}/\sigma^{\max}_{QC}$ for different values of $n$. 
Since, $p\in[0,1]$, instead of considering the usual Cauchy-Lorentz distribution, for the classical algorithm, we take half Cauchy distribution within the desired range, i.e., [0,1]. The corresponding distribution function is given by 
\begin{equation}
    f_{CC}(p)= \frac{N_{CC}}{1+ p^2 /v_C^2}
\label{eq-cauchy-classical} ,\nonumber
\end{equation}
where $N_{CC}$ is the normalization constant, found by considering the allowed range of $p$ as [0,1], given by $N_{CC}=1/(v_C \tan^{-1}(1/v_C))$. $v_C$ is a parameter which describes the spread of the distribution and can take any value within the range $0$ to $1$.
 The mean and the strength of the disorder distribution are given by
\begin{eqnarray}
    m_{C} &=&  \frac{N_{CC} v_C^2}{2} \ln(1 + \frac{1}{v_C^2})~\text{and}\nonumber\\
     \sigma_{CC} &=&  v_C \sqrt{N_{CC}( 1 - v_C \tan^{-1}(1/v_C))} .\nonumber
 \end{eqnarray}
 The highest strength attainable by this disorder is $\sigma_{CC}^{max}=1/\sqrt{3}$, i.e., when it becomes equivalent to the uniform distribution of $p$ over the full range [0,1].
The disorder-averaged probability of success, $C_C=(1-m_C)/2^{n-1}$, of the classical algorithms in a single query in the presence of Cauchy-Lorentz disorder is plotted against the scaled disorder strengths, $\Bar{\sigma}_C=\sigma_{QC}/\sigma^{\max}_{QC}$, in the same figure, i.e., Fig. \ref{fig:cauchy-fit}. From the plots, it is clear that whatever be the disorder strength even when $n=1$ the classical algorithm could not defeat its quantum variation. We also observed that the hierarchy of the quantum algorithm increases with $n$ at least up to $n=10$. For $n=10$, $C_C$ reduces to the order of $10^{-3}$ whereas the quantum algorithm still provides a significant probability of success, for weak disorder. If the strength of the disorder is increased above a moderate amount, $Q_C$ starts to decrease rapidly, finally becoming almost zero. It can be noticed from the figure that $Q_C$ exhibits a concave trend for the weak disorder but switches to a convex behavior at higher values of disorder strength, $\Bar{\sigma}_C$. This feature is more significant for $n=10$. in The profile of the curve, $Q_{C},$ can be fitted using the function 
$$Q_{C} = a_C \exp(-b_C  \Bar{\sigma}_{C} ^2) + c_C  \Bar{\sigma}_{C}^2 +e_C  ,$$
where $a_C$, $b_C$, $c_C$, and $e_C$ are the fitting parameters.
Values of all of these fitting parameters, $a_C$, $b_C$, $c_C$, and $e_C$, have been provided in Appendix  B.

\subsection{Discrete disorder}
Next, we try to compare the quantum and classical algorithms in presence of the discrete disorder distribution. 
The discrete circular disorder is considered to be a cousin of a discrete distribution in one dimensional parameter. The discrete disorder that we consider to be present in the classical algorithm takes fixed value $p=v_D$ with unit probability where $v_D$ defines the disorder strength.   
The mean of this distribution is 
\begin{equation}
    m_{D}={v_D}\nonumber
\end{equation}
and disorder's strength is also given by
\begin{equation*}
    \sigma_{CD} = v_D .
\end{equation*}
Similar to the previous disorders, in order to compare the efficiencies of quantum and classical algorithms, we divide the disorder strength, $\sigma_{CD}$ ($\sigma_{QD}$), of the disorder acting on the bit flip (Hadamard) operator by the maximum strength of the same, $\sigma_{CD}^{max}=1$ ($\sigma_{QD}^{max}=\pi$), to define a new quantifier of the disorder's intensity, $\Bar{\sigma}_D=\sigma_{CD}/\sigma_{CD}^{max}$ ($\Bar{\sigma}_D=\sigma_{QD}/\sigma_{QD}^{max}$).

\begin{figure}[H]
    \centering
    \includegraphics[width=8.5cm]{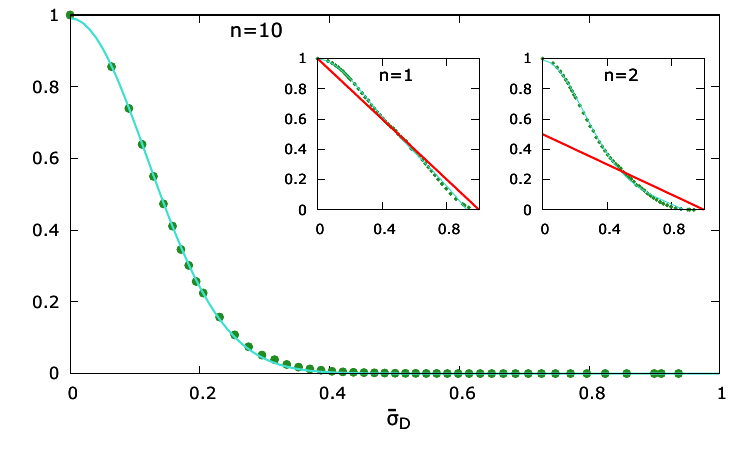}
    \caption{\textbf{Comparison between the efficiencies of the discrete disordered quantum and classical algorithms}. We represent behavior of $Q_D$ with respect to $\Bar{\sigma}_D$ using green points for $n=10$. The same has been plotted for $n=1$ and $n=2$ in the insets using the same color, green. To facilitate a comparison between the quantum algorithm and its classical counterpart, we also plot $C_D$ against $\Bar{\sigma}_D$ in the insets, using red points, for $n=1$ and 2. The sky-blue line represents the fitted curve for $n=10$.
    The axes are dimensionless.}
    \label{fig:circular-fit}
\end{figure}

In Fig. \ref{fig:circular-fit}, we plot the disorder-averaged probability, $Q_D$, of successfully guessing the string in a single query for discrete disordered BV algorithm, with respect to scaled disorder strength, $\Bar{\sigma}_D$, considering strings of different lengths. We also plot the disorder-averaged probability corresponding to the disordered classical algorithm, $C_D=(1-m_D)/2^{n-1}=(1-\Bar{\sigma}_D)/2^{n-1}$ for discrete disorder with respect to the same in Fig. \ref{fig:circular-fit}. For one bit noiseless case, the quantum and classical probabilities are the same, that is the secret string can be perfectly determined in both of the cases. Among all the considered types of disorders, only in the case of this discrete disorder, we see that the classical algorithm significantly outperforms the quantum one for at least $n=1$ and 2 when the strength of the disorder is much higher. We can notice from Fig. \ref{fig:circular-fit}, the quantum version is still more beneficial than the classical algorithm for weak or moderate disorders. In Sec.~\ref{al}, we will see that the difference between the capabilities of quantum and classical algorithms almost reduces to zero for all disorder strengths when the size of the secret string in hand is very large.
For larger values of $n$, say $n=10$, the classical success probability becomes almost zero for all disorder strengths whereas $Q_D$ is still non-zero for weak disorder strengths. The nature of $Q_D$ shows a concave to convex transition with respect to disorder strength, $\sigma_{QU}$. Similar to the case of spherical Gaussian disorder, we fit the profile of the curves, shown in Fig. \ref{fig:circular-fit}, using a function given by
$$Q_D = a_D \exp(- b_D   {\Bar{\sigma}_{D}}  ^2)+c_D {\Bar{\sigma}_{D}^2} + e_D .$$
The values of the fitting parameters, $a_D$, $b_D$, $c_D$ and $e_D$ are provided in Appendix B.


\subsection{Squeezed disorder}
\begin{figure}[H]
    \centering
    \includegraphics[width=8.5cm]{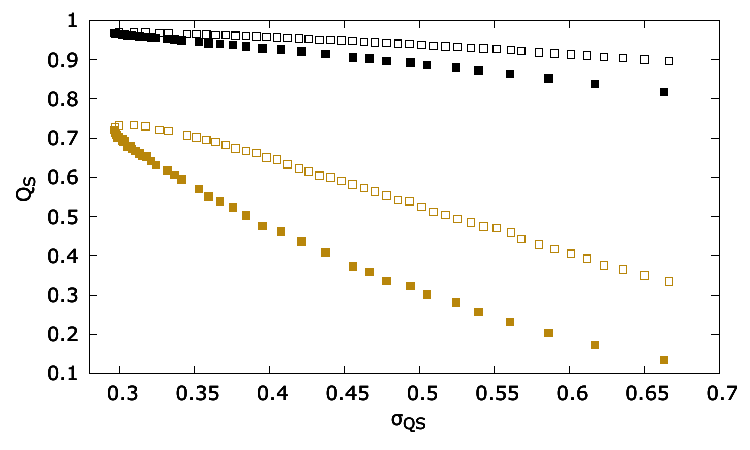}
    \caption{\textbf{Behavior of the squeezed disorder-averaged success probability of determining the correct secret string using BV algorithm}. We plot $Q_S$ along the vertical axis with respect to the disorder strength, $\sigma_{QS}$, which has been shown along the horizontal axis. We have considered $D=0.524$ (i.e. $\pi/6)$  which restricts the range of $r$ to $[0.17,6]$. The  hollow and solid squares represent the case when the disorder is considered to be squeezed along the $z$-axis ($r>1$) and $y$-axis ($r<1$), respectively. The length of the secret string is considered to be $n=1$ (black points) and $n=10$ (light-brown points). All the considered quantities are dimensionless.} 
    \label{fig-squ-prob}
\end{figure}

Lastly, we consider the squeezed distribution of disorder. As mentioned before, it is not rotationally symmetric about the mean. We are interested in visualizing how the success probability for BV algorithm varies with respect to the disorder strength of the squeezed disorder. Other than the mean, the disorder in this case is dependent on two more parameters: The spread of the disorder around the mean, and the amount of squeezing. We quantify the spread using the area of the distributed points' projection, $D$, and the squeezing is quantified using the ratio of the lengths of axes parallel to $y$-axis and $z$-axis, $r$. $r>1$ ($<1$) represents a distribution squeezed along 
$z$-axis ($y$-axis). We took $D=0.524$, varied $r$ within the allowed range, i.e., from $D/\pi$ to $\pi/D$, and calculated the corresponding strength of the disorder, $\sigma_{QS}$. In Fig. \ref{fig-squ-prob}, we plot the disorder-averaged probability of obtaining the correct string for BV algorithm in the presence of squeezed disorder, as a function of the disorder strength, $\sigma_{QS}$ considering $D=0.524$. Since $\sigma_{QS}$ does not depend on which direction it has been squeezed, the value of $\sigma_{QS}$ is the same for $r=h$ and $r=1/h$ for all $h$ for fixed $D$. But, as we can see from Fig. \ref{fig-squ-prob}, the disorder-averaged probability, $Q_{S}$, of correctly detecting the string of fixed size using BV algorithm does depend on the direction of squeezing. From the figure, it is clear that disorders with $r<1$ damage the algorithm more than $r>1$ disorder for fixed $n$.

It is not possible to incorporate any asymmetry in the classical algorithm's disorder, since the disorder, in this case, is only being incorporated inside one parameter, $p$. Thus, we are unable to compare the squeezed disordered BV algorithm with an analogous classical algorithm.

\begin{figure*}
\includegraphics[width=5.8cm]{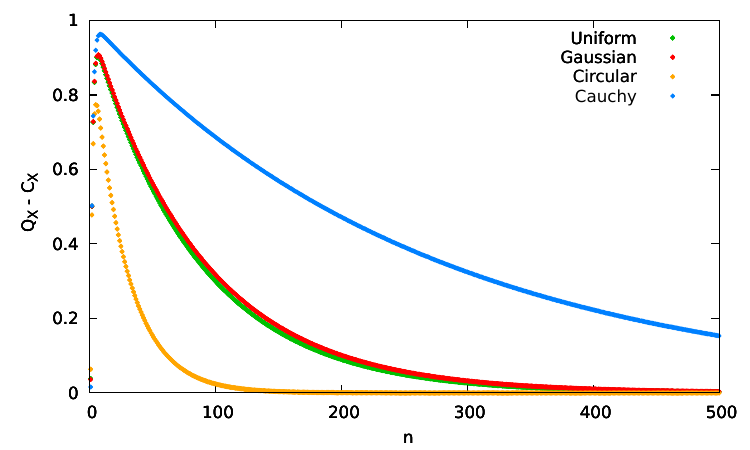}
\includegraphics[width=5.8cm]{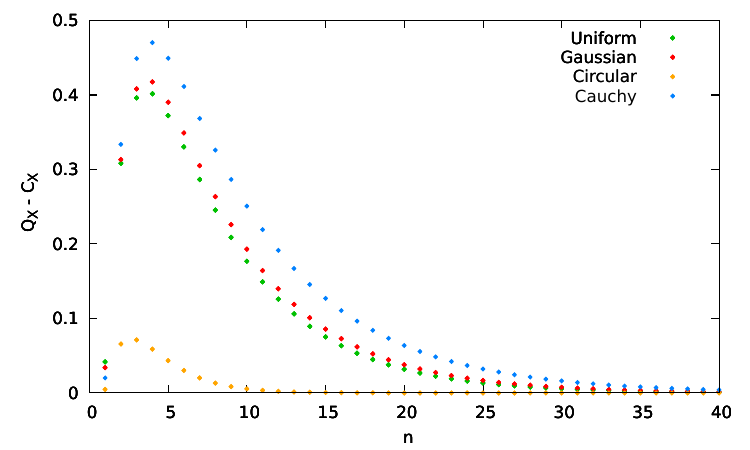}
\includegraphics[width=5.8cm]{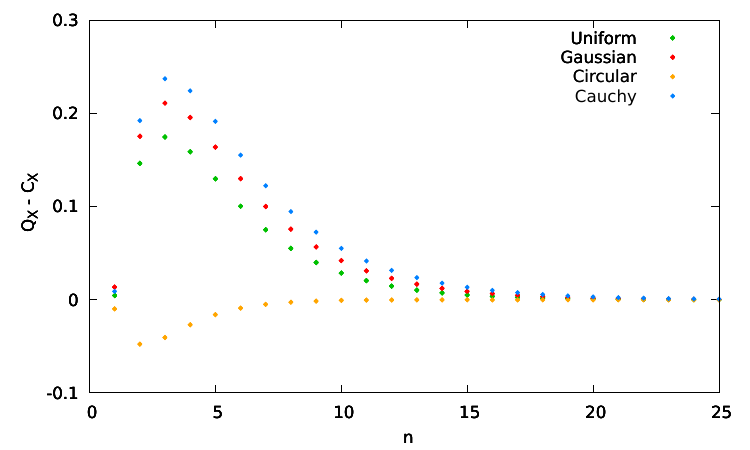}
\caption{\textbf{Difference in the efficiencies of the quantum and classical algorithms}. $Q_X-C_X$ is plotted by considering different types of disorder, $X$, with respect to the number of bits of the secret string. For example, for uniform disorder we plot $Q_U-C_U$, i.e., in this case, $X$ represents $U$. We consider four different types of noise, viz. uniform (green points), Gaussian (red points), Cauchy-Lorentz (blue points), and discrete (orange points). The disorder strength has been fixed at $0.1$, $0.4$ or $0.6$. The left plot represents the case when the disorder strength is $0.1$ whereas the plots in the middle and right correspond to disorder strengths of $\Bar{\sigma}_X=0.4$ and $0.6$ respectively. All the quantities are dimensionless.
}
 \label{fig:sig_fix-fit}
\end{figure*}


\subsection{Quantum vs. classical efficiency in presence of different disorders}
\label{al}
In this section, our primary objective is to examine the extent of quantum advantage in the BV algorithm over its classical counterpart. In this regard, we determine the difference in quantum and classical disorder-averaged success probabilities, $Q_X-C_X$, across varying numbers of bits, in the presence of different disorders, $X$, of fixed strength. In Fig.~\ref{fig:sig_fix-fit}, we present the behavior of $Q_X-C_X$ as a function of the length of the string, $n$. In particular, we consider the four types of disorder distributions, discussed in the previous sections, viz. uniform, Gaussian, Cauchy-Lorentz, and discrete.
From Fig.~\ref{fig:sig_fix-fit}, it can be concluded that for all considered disorders having small scaled strengths, as the length of the string is increased, the difference, $Q_X-C_X$, increases till it reaches a certain maximum value, after which it starts decreasing. When numerous bits are considered, this difference between the quantum and classical algorithms becomes negligible.  When the strength of the disorder is increased to a higher value, the qualitative behavior of $Q_X-C_X$ still remains the same for all the disorders, except the discrete disorder.
In the case of discrete disorder, when we consider $\Bar{\sigma}_X$ as high as 0.6, the plot of $Q_X-C_X$ depicts a drastic change to a rather opposite nature, the classical algorithm becoming more powerful than the BV algorithm for small values of $n$. The range of the string-length for which there exists a significant gap between the efficiencies of the two algorithms depends on the strength of the disorder. Specifically, if we increase the disorder strength the range decreases for all types of disorder we considered.


The peak in quantum advantage is observed when the number of bits in the secret string is small. This is due to the fact that, in the presence of a fixed amount of disorder, as the number of bits increases, the success probability of both quantum and classical algorithms decreases. But the efficiency of the classical algorithm reduces by a factor of $1/2^{n}$ even for the noiseless case. So, after a particular value of $n$, say 10, the classical success probability becomes almost zero for the complete range of disorder strength for all the disorders. On the other hand, the success probability of the disordered BV algorithm reduces slowly with $n$ for every disorder whatever its strengths. Thus, the difference between the quantum and classical success probabilities initially increases with $n$, until a certain value of $n$, beyond which it decreases as $n$ continues to increase.


\section{Conclusion}
\label{sec7}
BV algorithm determines the hidden bit string encoded inside an oracle with 100\% accuracy in a single query. However, in practical scenarios where quantum operations are affected by unwanted noise and disturbances, it becomes crucial to investigate their effects on the algorithm's performance. 

We incorporated various types of disorders inside the Hadamard gates which are used in the algorithm. The disorders force the Hadamard gate to rotate a state in random directions. By considering disorders having reflection symmetry, we found that the disorder-averaged success probability of correctly determining the string does not depend on the actual form of the string but only on the length of the string and the characteristics of the disorder. Using the central limit theorem we proved that if the disorder's distribution has a well-defined mean and variance and the number of bits required to be detected is large, then the effect of the disorder on the quantum BV algorithm would not depend on the type of the disorder but only on the mean and variance of the disorder. 

The noisy Hadamard gates project the state, $\ket{0}$, to another state on the Bloch sphere. That state has been chosen from different probability distributions. The distributions under consideration were Haar-uniform, spherical Gaussian, spherical Cauchy-Lorentz, discrete circular, and squeezed. We compared the performance of the noisy BV algorithm with an analogous noisy classical algorithm. 
We found that the quantum BV algorithm demonstrates its efficacy in terms of the probability of correctly guessing the secret string in a single query, when the size of the string is small or moderate and the strength of the disorder is not high, regardless of the type of disorder distribution. If we consider large secret strings, the success probability of the BV algorithm merges with the same of classical algorithm for all disorders of any strength. The range of the length of the string for which the disordered BV algorithm significantly overpowers disordered classical algorithm depends on the amount of incorporated disorder. For higher amount of disorder the range becomes narrower.
Precisely, we examined that though disordered BV algorithm is not efficient for very large string, it can still be effective for detecting strings having moderate or small lengths when the strength of the disorder is not very high.

\section*{Acknowledgment}
KS acknowledges support from the project MadQ-CM
(Madrid Quantum de la Comunidad de Madrid) funded
by the European Union (NextGenerationEU, PRTRC17.I1) and by the Comunidad de Madrid (Programa
de Acciones Complementarias).

\section*{Appendix A}
\label{A}
In the BV algorithm, the initial state of input qubits is considered to be $\ket{\psi_0} = \ket{0}^{\otimes n}$. After application of noisy Hadamard gates, $H(\theta_k,\phi_k)$ the state of the system of $n$-qubits becomes 

\begin{equation*}
\begin{split}
    \ket{\psi_1}  = &\bigotimes_{k=1}^n H(\theta_k,\phi_k) \ket{0}^{\otimes n}\\
     = &\left (\cos \frac{\theta_n}{2} \ket{0} + e^{i \phi_n} \sin \frac{\theta_n}{2} \ket{1} \right ) \otimes \\
    & \ldots \otimes \left (\cos \frac{\theta_1}{2} \ket{0} + e^{i \phi_1} \sin \frac{\theta_1}{2} \ket{1} \right ) \\
    = & \bigotimes_{k=1} ^n \left( \cos \frac{\theta_k}{2} \ket{0} + e^{i \phi_k} \sin \frac{\theta_k}{2} \ket{1} \right ) ,
    \end{split}
\end{equation*}
where $\theta_i$ and $\phi_i$ describe the independent local disorder, present in the noisy Hadamard gate $H_i$. Querying the oracle with state $\ket{\psi'_1}$ produces the state 
\begin{equation*}
    \ket{\psi_2}= \bigotimes_{k=1} ^n \left(\cos \frac{\theta_k}{2} \ket{0} + (-1)^{s_k} e^{i \phi_k} \sin \frac{\theta_k}{2} \ket{1} \right ).
\end{equation*}
Applying another set of $n$ independently noisy Hadamard gates on the state $\ket{\psi'_2}$ we get
\begin{eqnarray*}
    \ket{\psi_3}&=&\bigotimes_{k=1}^n H(\theta'_k,\phi'_k)\ket{\psi'_1}\\
    &= &\bigotimes_{k=1} ^ n  \Bigg[\cos \frac{\theta_k}{2} \left(\cos\frac{\theta_k ' }{2} \ket{0} + e^{i \phi_k ' } \sin\frac{\theta_k '}{2} \ket{1}  \right) \\ &
  + & (-1)^{s_k} e^{i \phi_k} \sin \frac{\theta_k}{2} \left(\sin \frac{\theta_k '}{2} \ket{0} - e^{i \phi_k ' } \cos \frac{\theta_k '}{2} \ket{1} \right)\Bigg].
\end{eqnarray*}
On simplifying, we get
\begin{align*}
&\ket{\psi_3}\\
&=\bigotimes_{k=1} ^ n  \Bigg[\left( \cos \frac{\theta_k}{2} \cos\frac{\theta_k ' }{2} + (-1)^{s_k} e^{i \phi_k } \sin\frac{\theta_k }{2} \sin\frac{\theta_k ' }{2}\right) \ket{0} \\ &+ e^{i \phi_k '} \left(\cos \frac{\theta_k}{2} \sin \frac{\theta_k '}{2} - (-1)^{s_k} e^{i \phi_k } \sin \frac{\theta_k}{2} \cos\frac{\theta_k ' }{2} \right) \ket{1}\Bigg].
\end{align*}


\section*{Appendix B}
\label{B}
This appendix provides values of the fitting parameters of all the fitted curves discussed in Sec. \ref{sec6}. 
\begin{enumerate}
    \item \textbf{Uniform Disorder with finite cut-off} : The success probability, $Q_U$, in presence of the uniform disorder are represented for different values of $n$ in Fig.~\ref{fig:uniform-fit}. The behavior of these curves can be determined using least-square fitting, with fitting parameters dependent on $n$. The profile of the curve of $Q_U$ vs $\Bar{\sigma}_U$ matches the following function
    $$Q_U = a_U \exp(-b_U {\Bar{\sigma}_{U}}^2)+e_U ,$$ 
   where $a_U$, $b_U$, and $e_U$ are the fitting parameters.
    \begin{itemize}
        \item For $n=1$, the fitting parameters are given as  $a_U=0.59 \pm 0.0024$, $b_U=1.8 \pm 0.018$, and $e_U=0.41 \pm 0.0027$ with an error of $0.0023$.

        \item  For $n=2$, the parameters become $a_U = 0.79 \pm 0.0027$, $b_U = 2.6 \pm 0.027$, $e_U = 0.19 \pm 0.0028$ with an error of $0.0046$.

        \item Finally, for $n=10$, the parameters become $a_U=0.99 \pm 0.0028$, $b_U=10 \pm 0.068$, $e_U = 0$ with an error of $0.0059$.
    \end{itemize}

    \item \textbf{ Gaussian Disorder:} The profile of the curves depicted in Fig.~\ref{fig:gaussian-fit} to represent the success probability of the BV algorithm can be well-fitted using the least square fitting technique.
    The function which appropriately fits the plot of $Q_G$ with respect to $\Bar{\sigma}_G$ is
     $$Q_G = a_G \exp (- b_G {\Bar{\sigma}_{G}^2} ) + c_G {\Bar{\sigma}_{G}} ^2 + e_G  ,$$ 
    
    \begin{itemize}
        \item  For $n=1$, the parameters are $a_G = 0.19 \pm 0.0031$, $b_G = 4.3 \pm 0.087$, $c_G = -0.31 \pm 0.0034$ and $e_G = 0.80 \pm 0.0033$ with an error of $0.0012$.

        \item For $n=2$, $a_G = 0.43 \pm 0.0044$, $b_G = 4.4 \pm 0.059$, $c_G = -0.33 \pm 0.0048$, and $e_G = 0.57 \pm 0.0046$ with an error of $0.0018$.

        \item  For $n=10$,  $a_G = 0.95 \pm 0.0031$, $b_G = 10 \pm 0.089$, $c_G = -0.052 \pm 0.0044$ and $e_G = 0.046 \pm 0.0029$ with an error of $0.0045$.
    \end{itemize}

 \item \textbf{Cauchy-Lorentz Disorder}
 In Fig.~\ref{fig:cauchy-fit}, disorder-averaged probabilities of success of BV algorithm is plotted against the scaled disorder strength, $\bar{\sigma}_C$. The curves drawn for different $n$ can be fitted using the least-square method. The following function can be used to fit the nature of $Q_C$ against $\Bar{\sigma}_C$
$$Q_{C} = a_C  \exp(-b_C  {\Bar{\sigma}_{C}}  ^2) + c_C  {\Bar{\sigma}_{C}} ^2 +e_C.$$ 

 \begin{itemize}
     \item  For $n=1$, $a_C=0.25 \pm 0.0035$, $b_C=2.5 \pm 0.032$, $c_C=-0.27 \pm 0.0028$, $e_C=0.75 \pm 0.0036$ with an error of $0.00056$.

     \item  For $n=2$, $a_C= 0.52 \pm 0.0052$, $b_C= 2.9 \pm 0.029$, $c_C= -0.26 \pm 0.0046$, $e_C=0.47 \pm 0.0053$, with an error of $0.0012$. 

     \item For $n=10$, $a_C= 0.95 \pm 0.0026$, $b_C= 9.3 \pm 0.054$, $c_C= -0.044 \pm 0.0037$, $e_C=0.030 \pm 0.0026$, with an error of $0.0044$. 
 \end{itemize}

\item \textbf{Discrete Circular disorder: } The behavior of the different curves shown in Fig.~\ref{fig:circular-fit} can be fitted using the least square method. The following function mimics the behavior of $Q_D$ with respect to $\Bar{\sigma}_D$
$$Q_D = a_D \exp(- b_D   {\Bar{\sigma}_{D}}  ^2)+c_D {\Bar{\sigma}_{D}^2} + e_D .$$
\begin{itemize}
    \item  For $n=1$, the parameters are given as $a_D = 0.46 \pm 0.029$, $b_D = 5.5 \pm 0.55$, $c_D =-0.58 \pm 0.040$, $e_D = 0.52 \pm 0.032$ with an error of $0.017$.

    \item  For $n=2$, the parameters are given as $a_D = 0.82 \pm 0.0015$, $b_D = 7.6 \pm 0.28$, $c_D = -0.20 \pm 0.023$, $e_D = 0.16 \pm 0.016$ with an error of $0.014$.

    \item  For $n=10$, the parameters are $a_D = 0.99 \pm 0.029$, $b_D = 35.4 \pm 0.17$, $c_D =0$, $e_D = 0$, with an error of $0.0035$.
\end{itemize}
\end{enumerate} 

\bibliography{bv}

\end{document}